\RequirePackage[T1]{fontenc}
\documentclass[conference]{IEEEtran}
\usepackage{cite}
\usepackage{amsmath,amssymb,amsfonts}
\usepackage{graphicx}
\usepackage{url}
\usepackage{balance}
\usepackage{xcolor}
\usepackage[hidelinks]{hyperref}

\begin{document}
\title{Path-Aware Fusion of Infrastructure RF Sensing and Robot Perception in a Live Robotic Digital Twin}
\newif\ifanon \anonfalse
\ifanon
\author{\IEEEauthorblockN{Anonymous Authors}\IEEEauthorblockA{Paper under double-blind review}}
\else
\author{\IEEEauthorblockN{Alberto Mart{\'i}nez M{\'e}ndez, Antonio de la Oliva, and Carlos J. Bernardos}
\IEEEauthorblockA{Universidad Carlos III de Madrid, Legan{\'e}s, Spain\\
alberto.martinez@uc3m.es, aoliva@it.uc3m.es, cjbc@it.uc3m.es}}
\fi
\maketitle

\begin{abstract}
Mobile robots in industrial workspaces must react to people that their onboard
sensors cannot yet observe. Communication infrastructure can fill part of this
gap by acting as a sensor, but its observations differ from camera and LiDAR
evidence in meaning, spatial support, and rate. We present a live robotic
Digital Twin (DT) that integrates heterogeneous sensing branches---a commodity
60-GHz link, joint Wi-Fi/5G sensing, a fixed camera, and robot-borne
LiDAR---as spatially grounded observations with explicit freshness, and that
uses the robot's live pose and heading to decide whether evidence lies on the
path the robot is about to sweep. We show that a 60-GHz link carrying
near-gigabit traffic detects human presence through the phase signature of
its antenna array (balanced accuracy 0.98 on unseen sessions and 0.999 from
one day to the next in a first site), but that a second site with a different
link geometry requires a site-specific model (0.98 on unseen sessions after
retraining, 0.73 without). On a Unitree Go2 patrolling under the DT, every
person entering the camera field or the 60-GHz link stopped the robot (20/20);
for the camera, the median time from physical entry to standstill was 1.83~s,
of which detection, DT update, and networking take about 0.15~s and the
rest is the 0.5-s confirmation and the robot's braking. Path-aware
gating removed all stops caused by off-path presence (5 stops versus 0 for the
same scenario), stale branches stopped the robot in fail-safe mode (6/6) and
were ignored otherwise (0/6), and five unattended laps produced no false stop.
\end{abstract}


\begin{figure*}[!t]
    \centering
    \includegraphics[width=\textwidth,trim=0 430bp 0 0,clip]{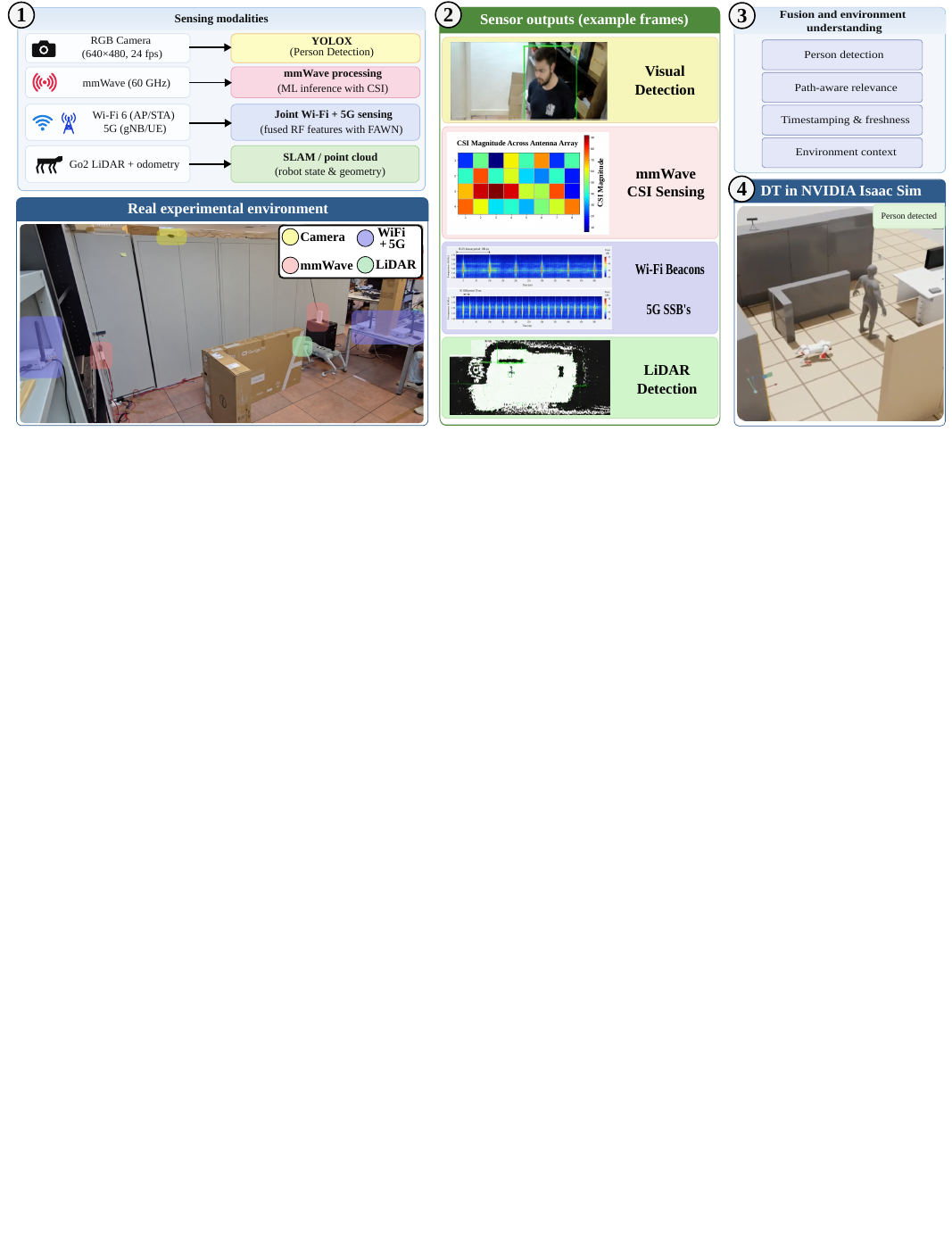}
    \caption{Heterogeneous sensing-to-DT architecture: sensing branches and
    testbed (panel~1), representative outputs (panel~2), fusion and environment
    understanding (panel~3), and the NVIDIA Isaac Sim DT (panel~4).
    Feature-level fusion takes place only inside the Wi-Fi/5G branch (FAWN);
    the DT integrates the outputs of all branches as timestamped, spatially
    grounded evidence for path-aware decisions.}
    \label{fig:system-overview}
\end{figure*}

\section{Introduction}
\label{sec:introduction}

Mobile robots increasingly share industrial workspaces with people and
equipment. Their onboard perception is inherently local: cameras depend on
field of view, illumination, and occlusion, whereas light detection and
ranging (LiDAR) measures geometry without identifying what it observes.
Relevant changes therefore often occur where the robot cannot yet see.
Digital Twins (DTs)~\cite{tao} offer a natural runtime representation in which
robot state and environmental information can be maintained jointly and
supervised remotely.

Integrated sensing and communications (ISAC)~\cite{isac} adds a complementary
source of evidence: links that are already deployed for connectivity also
respond to people and objects through their channel state information (CSI).
POLAR localizes passive objects with IEEE~802.11ad phased
arrays~\cite{polar}, waveSLAM combines commodity mmWave radios with LiDAR for
indoor mapping~\cite{waveslam}, MultiLoc and MILAGRO exploit sub-6-GHz and
mmWave observations jointly~\cite{multiloc,milagro}, and FAWN fuses Wi-Fi and
5G features for indoor scene inference~\cite{fawn}. The MultiX vision extends
this principle across technologies, bands, and external
sensors~\cite{multix}. On the DT side, Andrei \emph{et al.} integrate sensing,
communication, and robotics in a twinning platform~\cite{andrei}, Zhu
\emph{et al.} build a quadruped DT with multi-sensor scene
perception~\cite{zhu}, and the system in~\cite{previous} fed a live robotic
DT with 60-GHz CSI inference.
Infrastructure-assisted perception is established in vehicular settings,
where roadside cameras and LiDAR extend the field of view of vehicles in
cooperative 3D detection~\cite{dairv2x}, and multi-modal datasets pair
communication or RF sensing with cameras and LiDAR~\cite{mmfi,deepsense6g}.

These works either study one sensing family or fuse modalities into a single
estimate. Bringing infrastructure RF sensing into robot operation raises a
different problem. The modalities expose different \emph{semantics} (person,
blockage, occupied cell, geometric obstacle), different \emph{spatial supports}
(a camera frustum, a link segment, a floor cell, a region around the robot),
and different rates. The simplest integration---stopping whenever any branch
reports something---discards exactly this information. Assuming independent
false alarms across the $|\mathcal{S}|$ branches, the probability that at
least one of them falsely triggers this OR rule at a given decision instant is
$1-\prod_s(1-p_{\mathrm{fa},s})$, which grows with every added branch;
temporal confirmation lowers the resulting rate of false stops but does not
remove this growth. Every added sensor thus makes the robot \emph{less}
available, and the DT degenerates into a visualizer that plays no role in the
decision. Moreover, a robot that crosses an RF link is itself a
target for that link.

We argue that the DT should be the place where heterogeneous evidence is
related to what the robot is about to do. We build a live robotic DT around a
Unitree Go2 and NVIDIA Isaac Sim that ingests four branches---a commodity
60-GHz link, joint Wi-Fi/5G sensing with FAWN, a fixed camera, and onboard
LiDAR---as observations annotated with source, freshness, and spatial support.
The DT intersects this evidence with a corridor ahead of the robot's live pose
and triggers an action only when relevant evidence lies in it; evidence that
stops arriving is either treated as blocking (fail-safe) or ignored. The
action itself (stopping, waiting, re-routing, slowing down) depends on the
deployment; our experiments stop the robot and resume when the corridor
clears.
The system targets \emph{safety-aware supervision}; it is not a certified
safety function.

The contributions of this paper are:
\begin{itemize}
\interlinepenalty=10000
\item A \textbf{path-aware integration model} in which every observation
carries its source, timestamp, and spatial support, and the DT evaluates it
against the corridor the robot is about to sweep, together with an explicit
stop condition for stale branches, decoupled from the deployment-specific
action it triggers (Section~\ref{sec:fusion-gating}).
\item A \textbf{characterization of a commodity 60-GHz link as a human-presence
tripwire} while it carries near-gigabit traffic, in two sites: a standing
person lowers the coherence of the array phase signature, and an SVM on
attenuation and phase coherence generalizes across sessions and days where a
32-dimensional magnitude classifier learns session signatures; a different
link geometry, however, requires a site-specific model
(Section~\ref{sec:eval-mmwave}).
\item An \textbf{end-to-end implementation} on a physical Go2 and Isaac Sim
in which every branch publishes to the DT through a common sensor interface
(Section~\ref{sec:system}).
\item An \textbf{experimental evaluation on the robot} of the latency from a
person's physical entry to the robot's standstill, decomposed per stage, of
path-aware versus always-stop gating, of fail-safe and tolerant handling of
a lost branch, and of false stops over unattended laps
(Section~\ref{sec:evaluation}).
\end{itemize}

\section{System and Testbed}
\label{sec:system}

Fig.~\ref{fig:system-overview} summarizes the architecture. Four sensing
branches observe complementary portions of a laboratory workspace. Each branch
publishes a lightweight state record to the DT; the DT combines these records
with the live robot state, and the resulting decision is applied to the
physical robot.

\subsection{Workspaces}
\label{sec:architecture-scenario}

\begin{figure}[t]
    \centering
    \includegraphics[width=\columnwidth]{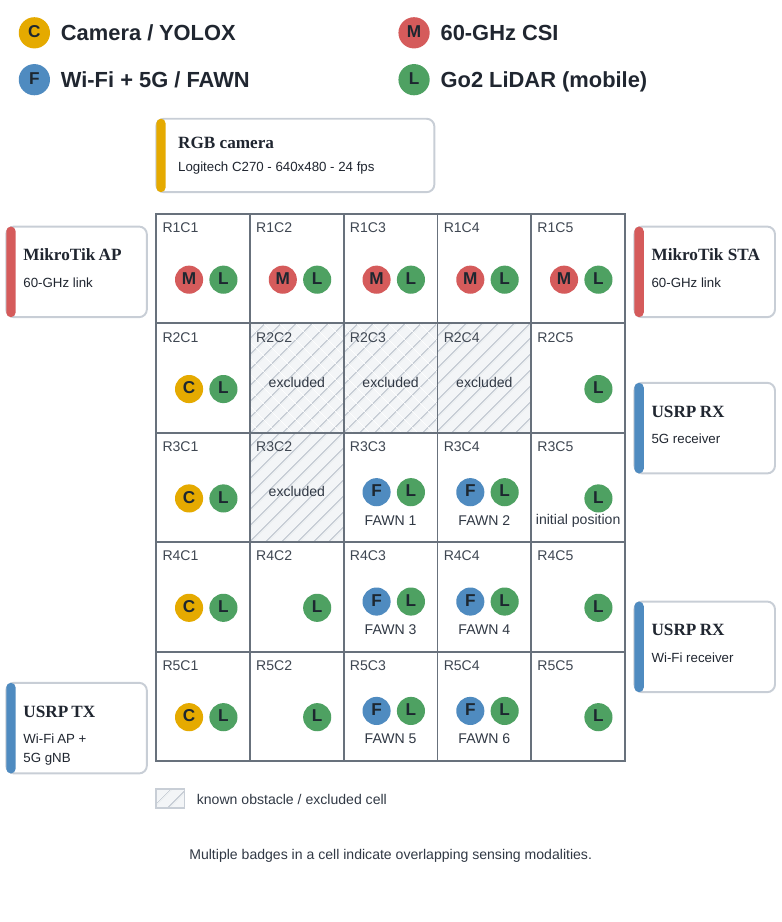}
    \caption{Spatial sensing coverage in site~A on the $60\times60$~cm grid.
    C, M, F, and L denote camera, 60-GHz link, joint Wi-Fi/5G, and mobile
    LiDAR; hatched tiles are excluded. The 60-GHz AP and STA face each other
    across row~1 (3~m).}
    \label{fig:spatial-coverage}
\end{figure}

We use two laboratories. The first (site~A) is the testbed in which the four
branches were deployed and the Wi-Fi/5G and first 60-GHz datasets were
collected: a $5\times5$ grid of $60\times60$~cm tiles
(Fig.~\ref{fig:spatial-coverage}) that serves as common spatial reference
$\mathcal{G}$. No branch covers the whole workspace: the camera observes the
corridor-facing region, the 60-GHz pair senses the region around its line of
sight (LoS), the Wi-Fi/5G branch resolves six predefined tiles, and the LiDAR
moves with the robot. The room is furnished, with metal cabinets behind the
60-GHz STA, i.e., a multipath-rich setting.

The online experiments run in a second laboratory (site~B). There the 60-GHz
AP and STA face each other 6.7~m apart at 1.0~m above the floor, the fixed
camera covers one side of the robot's patrol, and the robot patrols a square
loop. Site~B also provides a second 60-GHz dataset with a different link
geometry (Section~\ref{sec:offline-methodology}). The integration in
Section~\ref{sec:heterogeneous-fusion} only assumes that the supports of the
branches and the robot pose are known to the DT.

\subsection{Platform}
\label{sec:platform}

Table~\ref{tab:platform-config} lists the hardware and principal parameters.
The 60-GHz branch uses two MikroTik wAP~60G devices (AP/STA) running the IMDEA
Networks research OpenWrt firmware~\cite{mikrotiktools}, which exposes CSI
through the modified \texttt{wil6210} driver. The joint Wi-Fi/5G branch uses
USRP front ends transmitting 20-MHz Wi-Fi beacons and 5G Synchronization
Signal Blocks (SSBs)~\cite{fawn}. A fixed Logitech C270 provides the camera
stream, captured at $640\times480$; it delivers, and YOLOX processes,
$\approx$24~frames/s (median frame interval 40~ms). All processing runs on an NVIDIA DGX Spark (Ubuntu~24.04).

\begin{table}[t]
\caption{Experimental platform and runtime configuration.}
\label{tab:platform-config}
\centering
\footnotesize
\begin{tabular}{@{}p{0.15\columnwidth} p{0.25\columnwidth} p{0.50\columnwidth}@{}}
\hline
\textbf{Branch} & \textbf{Platform} & \textbf{Configuration} \\
\hline
Robot & Unitree Go2 EDU & ROS~2 gateway (Cyclone DDS, Zenoh bridge); square patrol at 0.22~m/s \\
DT & DGX Spark & REST sensor registry; Isaac Sim~5.1 \\
60~GHz & 2$\times$ wAP~60G & 60.48~GHz (ch.~2); 32 magnitude + 32 phase values per report; datasets: 36~Hz with saturating iperf3 TCP; online: 20~Hz with 200-Mb/s iperf3 \\
Wi-Fi & USRP & 2.412~GHz; 20-MHz beacon; 52 L-LTF CSI \\
5G & USRP & 3.54144~GHz; 15.36~MS/s; $240\times4$ SSB grid \\
Camera & Logitech C270 & $640\times480$ (30~fps requested), $\approx$24~frames/s captured and processed; YOLOX-S (COCO) \\
LiDAR & Go2 onboard & PointCloud2 $\rightarrow$ planar scan; SLAM Toolbox, 0.05-m map \\
\hline
\end{tabular}
\end{table}

\subsection{Robot and Digital-Twin Runtime}
\label{sec:architecture-dt}

The Go2 is reached through a ROS~2 gateway that exposes its state
(\texttt{sportmodestate}: position, velocity, yaw, and yaw rate) and accepts
velocity commands; the gateway is bridged to the DT host with Zenoh. Isaac
Sim~5.1~\cite{isaacsim} renders the physical pose in a reconstruction of the
laboratory. Robot state and environmental evidence use separate paths:
kinematics travel over ROS~2, whereas each sensing branch is registered in the
DT with its type and pose and updates its state through a REST interface
(Section~\ref{sec:fusion-common}), so that branches can evolve independently
behind a common interface.

A supervisory controller on the DT host replays the patrol (straight legs of
7~s at 0.22~m/s, i.e., about 1.5~m, separated by 95$^\circ$ right turns; one
lap takes about 50~s), polls the DT at 20~Hz, and anchors the gateway's
odometry to the DT frame at the start of each run. The patrol is open loop;
between runs the robot is returned to its start mark. The corridor of
Section~\ref{sec:fusion-gating} is placed with this odometry, whose error we
did not measure against an external reference. A lateral error $e_y$ and a
heading error $e_\theta$ shift the corridor at the look-ahead distance by about
$e_y+D\sin e_\theta$, e.g., 0.17~m for $e_\theta=5^\circ$ at $D=2$~m. The
corridor half-widths we use (0.45 and 0.55~m) exceed the Go2 half-width
(0.16~m) by 0.29--0.39~m, so a point that the robot body would actually sweep
remains relevant while the shift stays below that margin; evidence near the
corridor edge can be misclassified in either direction.

\section{Sensing Branches and Path-Aware Integration}
\label{sec:heterogeneous-fusion}

\subsection{Spatially Grounded Observations}
\label{sec:fusion-common}

Let $\mathcal{S}=\{M,F,C,L\}$ denote the 60-GHz, Wi-Fi/5G, camera, and LiDAR
branches. Each branch publishes observations
\begin{equation}
    o_s(t)=\big(y_s(t),\,\tau_s,\,\Omega_s(t)\big),\qquad s\in\mathcal{S},
    \label{eq:sensor-observation}
\end{equation}
where $y_s\in\{0,1\}$ indicates positive evidence (a person or an obstacle),
$\tau_s$ is the time at which the branch produced the observation (frame
capture for the camera, processing of the CSI report for the 60-GHz link), and
$\Omega_s$ is the region to which the evidence applies. Branches keep their
own semantics; what they share is \emph{where} their evidence is valid. In the
implementation, each branch is a sensor registered in the DT, and every update
carries $y_s$, $\tau_s$, a monotonically increasing sequence number, and a
representative location $p_s$ of its support, configured per sensor at
registration (the point at which the DT renders the person). \emph{Branch
liveness} is monitored on the DT host from the progression of the sequence
number, with a timeout $T_{\max}=2$~s: a branch whose sequence number does not
advance for $T_{\max}$ is stale. This detects stalled publishers without
requiring synchronized sensor clocks. Since our branches increment the
sequence number only after processing a new camera frame or CSI report, a
stalled source also stalls the sequence; the mechanism does not, however,
bound the age of the measurement itself (e.g., data buffered upstream of the
branch). In \emph{fail-safe} mode a stale or missing branch counts as blocking
evidence; in \emph{tolerant} mode it is ignored and its region is left to the
remaining branches, which is the weaker guarantee. Modality-specific confidences are not used by the
policy.

\subsection{60-GHz Link as a Human-Presence Tripwire}
\label{sec:fusion-mmwave}

The STA requests CSI reports from the \texttt{wil6210} vendor interface while
the link carries TCP traffic, so sensing is obtained from a link that is
simultaneously in use for communication. Each report contains $N=32$
magnitudes $a_n$ and phases $\phi_n$ measured across the elements of the
phased array. In site~A the AP and STA face each other 3~m apart at 0.9~m
above the floor, so the first Fresnel zone has a radius of at most 6~cm at
mid-link: a standing person intersects it, whereas the Go2 (about 0.4~m tall)
passes 0.5~m below it. In site~B the link is 6.7~m long at 1.0~m height
(Fresnel radius $\le$9~cm).

Two physically motivated features summarize each report with respect to an
empty-room reference. The first is the
attenuation of the mean magnitude $\bar a(t)=\frac{1}{N}\sum_n a_n(t)$,
\begin{equation}
    \Delta(t)=20\log_{10}\frac{\bar a(t)}{\bar a_{\mathrm{ref}}}.
    \label{eq:mmwave-detector}
\end{equation}
The second is the coherence of the array phase signature with that of the
empty room. Phases are taken relative to element~0, which removes the common
random phase offset of each report, and compared with the circular mean
$\psi_k$ of $\phi_k-\phi_0$ in the reference:
\begin{equation}
    c(t)=\Big|\frac{1}{N-1}\sum_{k=1}^{N-1}e^{\,j(\phi_k(t)-\phi_0(t)-\psi_k)}\Big|\in[0,1].
    \label{eq:mmwave-coherence}
\end{equation}
A change that scales the received signal uniformly across the array lowers
$\Delta$ but keeps $c$ close to its empty-room value, whereas one that alters
the relative phases across the array, e.g., by blocking the dominant path and
leaving other propagation paths, drives $c$ towards the value of uncorrelated
phases. An RBF-SVM~\cite{cortesvapnik} maps $(\Delta,c)$ to
\emph{human present} or \emph{absent}. For online operation the reference
is re-recorded from 20~s of empty room whenever the branch starts; the offline
generalization experiments use the fixed references of
Section~\ref{sec:offline-methodology}.

The per-report SVM decision is filtered in time by one of two alternatives,
both applied to the raw decisions. The \emph{stabilizer} accepts a new state
after two consecutive reports or three of the last five. The \emph{1-s vote}
keeps the raw decisions of the last second ($\approx$20 reports at 20~Hz) and
switches on when at least 70\% of them are positive and off when at most
30\% are; in between it keeps its previous state. The vote is the filter used
online, i.e., its output is the state published to the DT; offline we report
both. The vote suppresses sub-second fluctuations at the cost of delay: a
person must be seen by about 14 consecutive reports before the state turns
on. The support $\Omega_M$ is the region around the AP--STA segment.

\subsection{Joint Wi-Fi/5G Sensing}
\label{sec:fusion-fawn}

The second RF branch uses FAWN~\cite{fawn}: Wi-Fi CSI from 52 L-LTF
subcarriers and the $240\times4$ 5G SSB channel grid are canonicalized
(log-magnitude, common phase and linear phase-slope removal), encoded by
modality-specific convolutional--Transformer encoders, and fused by a
Transformer block. The output is person presence and, when occupied, one of
six tiles $\hat r_F$; the model was trained and evaluated in site~A
on the same six tiles~\cite{fawn}. We set $\Omega_F$ to the $3\times3$
block of tiles centred on $\hat r_F$ (the predicted tile and its eight
neighbours, diagonals included), or to the six-tile region when only presence
is reported. This one-tile margin is a conservative design choice that covers
confusions between adjacent tiles; it was not tuned on FAWN's confusion
matrix, and the FAWN branch does not take part in the online runs of
Section~\ref{sec:online-methodology}.

\subsection{Camera and LiDAR}
\label{sec:fusion-camera}

\textit{Camera:} YOLOX-S~\cite{yolox} with COCO weights~\cite{coco} runs on
every frame of the fixed camera; only \emph{person} detections with confidence
$\ge0.85$ and an upright box (height at least 1.15 times its width and at
least 15\% of the image height, which rejects the Go2) are kept, and the
decision passes through the same stabilizer as the 60-GHz branch, without
vote. Since the camera is fixed, its support $\Omega_C$ is the
floor region inside its field of view.

\textit{LiDAR:} the Go2 point cloud is projected to a planar scan and
registered in the SLAM map~\cite{slamtoolbox}. Points in previously free space,
farther than 0.22~m from the static baseline and 0.25~m from inflated walls,
are clustered (radius 0.45~m); a cluster of at least six points persisting for
three updates yields $y_L=1$, and $\Omega_L$ is the tile set covered by the
cluster. This branch reports geometric obstacles, not persons.

\subsection{Path-Aware Decision and Action}
\label{sec:fusion-gating}

We separate \emph{whether} the evidence affects the robot from \emph{what} the
robot does about it. The first question is answered by the DT, the second by a
use-case-specific action policy.

\textit{Baseline (global OR).} The simplest integration flags the robot
whenever any fresh branch is positive,
\begin{equation}
    z_{\mathrm{OR}}(t)=\textstyle\bigvee_{s\in\mathcal{S}}
    \big[y_s(t)\wedge \mathrm{fresh}_s(t)\big].
    \label{eq:semantic-or}
\end{equation}
It ignores $\Omega_s$: a person in any monitored region interrupts the robot.

\textit{Path relevance.} Let $\Pi(t)$ be the region the robot footprint will
sweep over a look-ahead distance $D$. We use a corridor attached to the live
robot pose $(x,y,\theta)$: a point lies in $\Pi(t)$ if its longitudinal
coordinate in the robot frame is in $[-b,D]$ and its lateral coordinate is
within $\pm w$. Evidence is relevant only if its support reaches that
corridor, $\Omega_s\cap\Pi(t)\neq\emptyset$:
\begin{equation}
\begin{split}
    z_{\mathrm{PA}}(t)=\textstyle\bigvee_{s\in\{M,F,C\}}
    \big[&y_s(t)\wedge \mathrm{fresh}_s(t)\\
    &\wedge\big(\Omega_s(t)\cap\Pi(t)\neq\emptyset\big)\wedge\neg\,\sigma_s(t)\big].
\end{split}
    \label{eq:path-aware}
\end{equation}
The look-ahead must cover the distance traveled before the robot is at rest,
$D\ge v\,T_{\mathrm{stop}}+r_R$, where $T_{\mathrm{stop}}$ is the time from the
person's entry to standstill (sensing, DT, confirmation, and braking) and
$r_R$ the footprint radius. Because the corridor follows the measured pose
and heading, it turns with the robot at corners and needs no route map.

\textit{Point approximation.} The evaluated implementation does not intersect
regions: it tests whether the configured point $p_s$ lies in $\Pi(t)$. This
approximates~\eqref{eq:path-aware} only when $\Omega_s$ is small relative to
the corridor. For an extended support, such as the region around a 6.7-m
link, part of $\Omega_s$ may enter the corridor while $p_s$ stays outside,
which is a spatial false negative. Testing the full support (a segment or
polygon against the corridor rectangle) removes it at the cost of more stops.
In the experiments of Section~\ref{sec:evaluation}, $p_s$ is therefore a
configuration of the scenario rather than an estimate of the person's
position.

\textit{Onboard safety layer.} The LiDAR branch is deliberately kept outside
this gating. Its support moves with the robot and always lies on the path, and
it does not depend on infrastructure that may fail, be stale, or be
mislocalized; an obstacle within the stop distance in front of the robot
therefore halts it regardless of $z_{\mathrm{PA}}$, acting as a last-resort
layer under the path-aware policy.

\textit{Self-explained evidence.} An infrastructure RF branch also responds to
the robot. The term $\sigma_s(t)$ is set for $s\in\{M,F\}$ when the robot
footprint at time $t$, known from the DT, intersects $\Omega_s$; a detection
of that branch is then attributed to the robot and ignored. The same
detection with the robot elsewhere is treated as a person. The price is a
blind interval for that branch while the robot crosses its support, during
which the onboard LiDAR covers its immediate surroundings. The suppression
deliberately masks the branch while the robot overlaps its support: it does
not distinguish robot-induced detections from simultaneous human presence, so
a person crossing the link together with the robot is also masked. In site~B
the patrol does not cross the link, $\sigma_M$ is never active, and the
mechanism is not evaluated.

\textit{Action.} When $z_{\mathrm{PA}}$ is confirmed, an action
$a\in\mathcal{A}$ is selected from the evidence held in the DT. Which actions
are available, and at what cost, is a property of the deployment rather than of
the sensing architecture: an industrial vehicle may \emph{wait} for the
region to clear, \emph{slow down} while passing a monitored but distant
detection, or \emph{re-route} through an alternative aisle when the expected
waiting time exceeds the extra travel time or energy of the detour, and the
remote operator may be notified in every case. The DT supplies what these
choices need---which regions are blocked, by what, and since when---through
the same interface. In our experiments the action is to \emph{stop} and to
resume the patrol where it was interrupted once the evidence clears.

\textit{Complete stop condition.} Eq.~\eqref{eq:path-aware} covers only
fresh evidence. The controller blocks the robot when any of the following
holds:
\begin{equation}
    z_{\mathrm{stop}}(t)=z_{\mathrm{pol}}(t)\vee z_{L}(t)\vee z_{\mathrm{stale}}(t)\vee z_{R}(t),
    \label{eq:stop}
\end{equation}
where $z_{\mathrm{pol}}$ is $z_{\mathrm{OR}}$ or $z_{\mathrm{PA}}$ depending
on the policy, $z_L$ is the LiDAR safety layer when that branch is enabled,
$z_{\mathrm{stale}}$ is set in fail-safe mode when an enabled branch is stale,
offline, or not registered, or the DT does not answer, and $z_R$ is set when
the robot state is older than 1~s. The operator can additionally hold the
robot.

\textit{Temporal hysteresis.} The blocking condition~\eqref{eq:stop} must
persist for $T_{\mathrm{occ}}=0.5$~s before the robot is stopped, and a clear
state must last $T_{\mathrm{clear}}=1.5$~s before the patrol resumes; the
controller starts blocked. The DT displays all fresh evidence regardless of the decision,
so the operator still sees off-path detections.

\section{Experimental Methodology}
\label{sec:experimental-methodology}

\subsection{Offline Branch Evaluation}
\label{sec:offline-methodology}

\textit{60-GHz link.} Each capture lasts 120~s at $\approx$36~Hz while
iperf3 saturates the link. In site~A the conditions are \emph{empty}, one
\emph{person} standing on the LoS centre, \emph{two persons} at 1/3 and 2/3 of
the link, and a \emph{metal plate} centred on the LoS at 0.9~m; they are
interleaved rather than recorded in blocks, so that condition and acquisition
time are not confounded, and repeated on two days four days apart with the
same antenna geometry: 10 sessions on day~1 and 8 on day~2 (two per
condition), 77\,864 reports in total. Both features use a single empty
reference recorded at the start of day~1, which is never used for testing.
In site~B we record 10 sessions (five empty, five with one person on the LoS
centre) in ABBA order, 43\,992 reports, with the first empty session as
reference. We evaluate leave-one-session-out (train on all sessions but one,
test on the held-out one) and across days (train on one day, test on the
other). Since consecutive reports are strongly correlated, random frame-level
splits are not used.

\textit{Wi-Fi/5G.} The FAWN model is trained and tested with session-level
60/20/20 partitions on captures with a person on each of the six tiles of
site~A and with the empty room; we refer to~\cite{fawn} for its results.

\subsection{Online Experiments}
\label{sec:online-methodology}

All four branches run online in the system; the quantitative runs below,
in site~B, use the fixed camera and the 60-GHz link as decision sources, which
isolates the infrastructure path that the robot cannot observe itself. The
60-GHz branch uses the site-B model. The robot's own front camera, also
registered in the DT, is logged but excluded from the decision, because in the
empty laboratory it flagged a person in 37\% of the samples. Each experiment
is a run of the patrol (Section~\ref{sec:architecture-dt}) with:

\textit{E1 (latency):} a person enters the camera field (10 repetitions) or
the 60-GHz link (10), stays about 5~s, and leaves; always-stop policy.

\textit{E2 (path relevance):} the same 60-GHz presence, published at a point
inside the square loop but off the robot's path, under the always-stop
policy~\eqref{eq:semantic-or} and the path-aware
policy~\eqref{eq:path-aware}; and, as a positive control, a person in the
camera field on the robot's path under the path-aware policy (6 repetitions
each). In E2 the corridor is $D=0.7$~m, $w=0.45$~m, $b=0.2$~m.

\textit{E3 (lost branch):} the camera or the 60-GHz branch is stopped three
times during the patrol and restarted, in fail-safe and in tolerant mode.

\textit{E4 (false stops):} five laps with nobody in the laboratory.

The controller logs, on the DT host's clock, every decision with its reasons
and the state of each branch, a DT snapshot at 10~Hz, the robot state at
20~Hz, and operator marks. The camera branch records its video with a
timestamp per frame; the physical entry of E1 is the first frame in which any
part of the person appears in the image, annotated by inspection
($\pm$1~frame, 42~ms). There is no video of the 60-GHz link, so its latency is
measured from the published detection. The robot is \emph{at rest} when its
speed is below 0.03~m/s and its yaw rate below 0.05~rad/s for 0.3~s; latencies
to standstill are taken at the start of that interval. Operator key presses
are used only to identify repetitions: their offset to the detection varied
by $\pm$0.3~s, so they are not used as ground truth.

\section{Experimental Evaluation}
\label{sec:evaluation}

\subsection{60-GHz Link: Human Presence Under Traffic}
\label{sec:eval-mmwave}

\textit{Coexistence.} CSI was obtained at 35.5--36.7~Hz with
$\ge$99.98\% request success in all 18 sessions while iperf3 saturated the
link. Goodput was 935--962~Mb/s in 14 sessions, bounded by the Gigabit
Ethernet port of the wAP~60G (PHY rate 2.31~Gb/s), so sensing does not
disturb a saturated wired-limited flow, although this does not measure the
residual over-the-air capacity. Four sessions of day~1 delivered
614--833~Mb/s; the drop did not recur on day~2 under any condition, so we do
not attribute it to the scene. In site~B, ten sessions ran at 36.4--37.0~Hz
($\ge$99.98\% success) with 773--969~Mb/s of goodput. During the online
experiments the iperf3 flow was capped at 200~Mb/s and CSI requested at 20~Hz
to bound the load on the DT host.

\begin{figure}[t]
    \centering
    \includegraphics[width=\columnwidth]{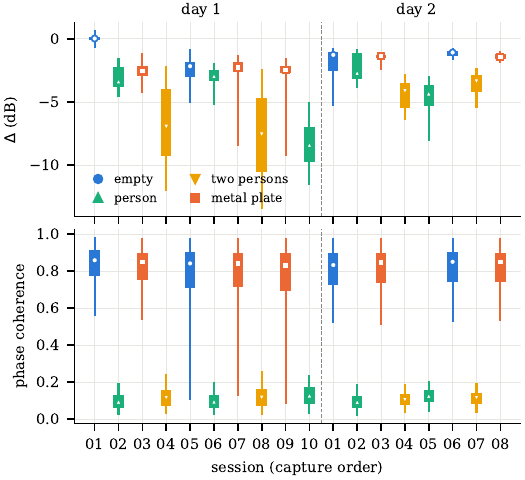}
    \caption{60-GHz features per session on two days under saturating traffic
    (box: interquartile range, whisker: 5th--95th percentile), both relative
    to one empty reference recorded at the start of day~1:
    attenuation $\Delta$~\eqref{eq:mmwave-detector} and phase-signature
    coherence $c$~\eqref{eq:mmwave-coherence}. With one or two persons $c$ drops
    to the level of uncorrelated phases; the metal plate, at the tested
    position, leaves it at its empty-room value.}
    \label{fig:mmwave}
\end{figure}

\textit{What the link detects.} Fig.~\ref{fig:mmwave} shows both features per
session. Every empty and metal-plate session keeps a median phase coherence of
0.83--0.86 with the day-1 reference, whereas every session with one or two
persons drops to 0.09--0.13, the level of uncorrelated phases (0.15 for 31
random phases). The signature recorded on day~1 still holds on day~2. The
magnitude is not decisive: the plate attenuates the link by 0.1--2.5~dB, one
person by 1.4--8.4~dB, and the empty link itself drifts by 1.1--1.3~dB
between days, so magnitude can neither tell a person from the plate nor count
persons.

\textit{Session- and day-disjoint detection.} With leave-one-session-out
evaluation over the 17 non-reference sessions, the RBF-SVM on $(\Delta,c)$
detects human presence with a balanced accuracy of 0.98 (TPR 99.9\%, FPR
3.5\% after stabilization). Trained on day~1 and tested on day~2, it reaches
0.999 (TPR 100\%, FPR 0.2\%); in the reverse direction 0.96 (TPR 99.7\%,
FPR 6.7\%), where the false positives are short episodes of at most 4~s in three
non-human sessions of day~1. The same SVM on $\Delta$ alone reaches 0.67--0.83
across protocols. After stabilization, a fixed threshold $c<0.4$ without any training
performs within 0.001 of the SVM in every protocol, and re-calibrating the reference
at the start of day~2 brings no gain: the decision rests on the phase
signature, which needs one empty capture rather than a labelled dataset.

\textit{Why not a 32-dimensional classifier.} An RBF-SVM on the 32 magnitudes
reaches 0.99 balanced accuracy on four classes when train and test come from
the same sessions of day~1, and it separates two sessions of the \emph{same}
condition with 0.90--0.99 accuracy, i.e., it learns session signatures; the
relative phase does not carry such a signature for static scenes
(0.49--0.56). For human presence, the 32-magnitude SVM drops to 0.95 (TPR
93\%) leave-one-session-out and to 0.91 (TPR 82\%) from day~1 to day~2.
Within-session accuracy therefore overstates what a CSI classifier learns about
the scene; all results above are session-disjoint.

\textit{A second site.} In site~B a person collapses the phase signature far
less (median $c$ of 0.37--0.60 in the five person sessions, against
0.83--0.85 empty) but attenuates the 6.7-m link more ($\Delta$ of $-3.3$ to
$-8.8$~dB). The site-A model, applied with a site-B empty reference, detects
only 47\% of the person reports (balanced accuracy 0.73), and the threshold
$c<0.4$ only 35\%. Retrained on site~B, the SVM reaches 0.95 leave-one-session
out after stabilization (TPR 92.4\%, FPR 3.1\%) and 0.98 with the 1-s vote
that feeds the DT (TPR 96.0\%, FPR 0.6\%); the weakest session is the one with
the least attenuation (detected 83\% of the time). Pooling both sites does
not produce a model that transfers between them (0.77--0.97, with up to 19\%
false positives). The phase signature therefore generalizes across sessions
and days for a given link geometry, but a deployment needs a short labelled
capture per site.

\subsection{End-to-End Latency}
\label{sec:evaluation-end-to-end}

\begin{figure}[t]
    \centering
    \includegraphics[width=\columnwidth]{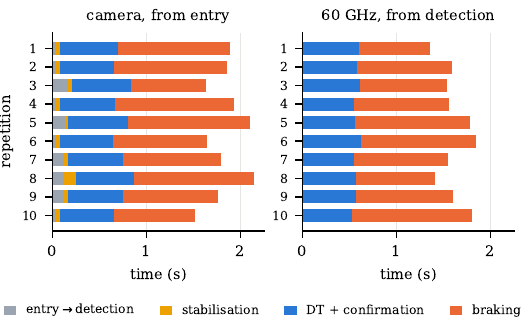}
    \caption{E1: time to standstill per repetition. Camera: from the person's
    first video frame; 60~GHz: from the detection published to the DT (no
    video of the link). DT + confirmation includes the PUT, the DT poll, and
    the 0.5-s confirmation.}
    \label{fig:latency}
\end{figure}

All 20 entries stopped the robot (Table~\ref{tab:online}). Fig.~\ref{fig:latency}
decomposes the delay. For the camera, the person is detected 82~ms (median;
39--163~ms) after appearing in the image, the detection is stable 44~ms later,
and the DT holds it 23~ms after that; the stop decision follows 587~ms after
the stable detection, i.e., the 0.5-s confirmation plus the DT poll. From
entry, the decision takes 723~ms (645--871~ms) and the robot is at rest after
1.83~s (1.51--2.15~s): braking from 0.22~m/s takes 1.12~s (0.80--1.29~s) and
dominates. For the 60-GHz link, the decision follows the published detection
after 573~ms (523--624~ms) and the robot is at rest after 1.57~s
(1.36--1.85~s); the time from the person cutting the link to the published
detection is not observable without video. It is dominated by the 1-s vote:
replaying the runtime vote at 20~Hz with the per-report detection rate of
site~B (91\%), the state turns on a median 0.74~s after the first CSI report
that sees the person (0.68--0.81~s for rates between 100\% and 83\%),
against 0.03~ms for features and SVM per report. The
robot travels about 0.40~m from a person's entry to standstill, which bounds
the look-ahead $D$ of the corridor from below. Resuming takes 1.57~s from a
clear corridor (the 1.5-s $T_{\mathrm{clear}}$) and the robot moves 29~ms
after the decision. In one 60-GHz repetition the detection dropped while the
person was still on the link: the robot resumed and was stopped again,
hence 11 stop decisions for 10 entries.

\subsection{Path Relevance}
\label{sec:eval-policy}

\begin{figure}[t]
    \centering
    \includegraphics[width=\columnwidth]{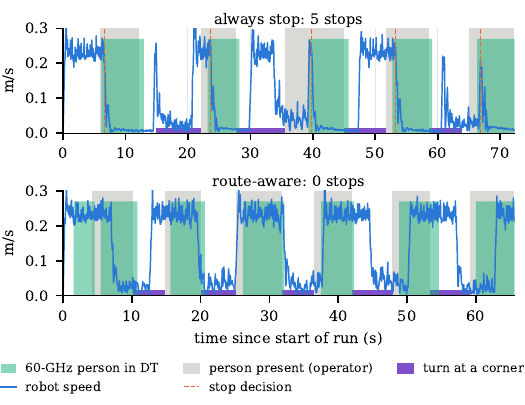}
    \caption{E2: the same off-path 60-GHz presence under the always-stop and
    the path-aware policy. The evidence reaches the DT in both runs; only the
    always-stop policy halts the robot. Speed dips without a stop decision are
    the turns at the corners of the loop.}
    \label{fig:e2}
\end{figure}

Fig.~\ref{fig:e2} shows the two E2 runs with the person inside the loop but
off the robot's path. The 60-GHz evidence reached the DT in every repetition
of both runs. Under the always-stop policy each of the five complete entries
stopped the robot (the first repetition was aborted by the operator before
the person entered). Under the path-aware policy none did: all 284 DT
snapshots with the person present were classified as off the corridor, and the
robot completed its lap. In the positive control, a person on the robot's path
in the camera field, the camera detected all six entries; the robot stopped
in the three repetitions in which its corridor reached the person (with
$D=0.7$~m) and correctly continued in the other three, when it was on another
side of the loop. One stop occurred 0.25~s after the person had started to
leave, i.e., the corridor reached a presence that the DT still held. The
decision thus depends jointly on the detection, the support held in the DT,
and the robot's pose and heading; with $D=0.7$~m the margin over the 0.40~m
traveled to standstill is small, and $D=2$~m is the value used in the
remaining runs.

\subsection{Lost Branches and False Stops}
\label{sec:eval-faults}

\begin{table}[t]
\caption{Online experiments in site~B (camera C, 60-GHz link M).}
\label{tab:online}
\centering
\footnotesize
\begin{tabular}{@{}p{0.42\columnwidth} c c r@{}}
\hline
\textbf{Experiment} & \textbf{Policy} & \textbf{Events} & \textbf{Stops} \\
\hline
E1 person enters, C & always & 10 & 10 \\
E1 person enters, M & always & 10 & 11$^{a}$ \\
E2 person off path, M & always & 5$^{b}$ & 5 \\
E2 person off path, M & path-aware & 6 & 0 \\
E2 person on path, C & path-aware & 6 & 3$^{c}$ \\
E3 C lost, fail-safe & always & 3 & 3 \\
E3 C lost, tolerant & always & 3 & 0 (+2 FP M)$^{d}$ \\
E3 M lost, fail-safe & always & 3 & 3 \\
E3 M lost, tolerant & always & 3 & 0 \\
E4 nobody, 5 laps & always & -- & 0 \\
\hline
\end{tabular}

\vspace{2pt}
\parbox{\columnwidth}{\footnotesize $^{a}$One premature resume and re-stop.
$^{b}$First of six repetitions aborted before entry. $^{c}$Stops only when the corridor reached
the person. $^{d}$Two 60-GHz false positives during a camera outage.}
\end{table}

With fail-safe handling, every outage stopped the robot (6/6): the branch
became stale 2~s after its last update, the decision followed after the 0.5-s
confirmation, and the robot was at rest 1.13~s (camera) and 1.30~s (60~GHz)
later (medians). When the branch came back, the patrol resumed 1.50--1.55~s
after the first fresh update. In tolerant mode no outage stopped the robot
(0/6). During one camera outage, however, the 60-GHz branch produced two false
positives that stopped the robot for 2.6 and 3.2~s: tolerating a lost branch
leaves the decision to the remaining ones, including their errors. We did not
determine whether the approaching robot caused them. Over five unattended laps
(251.6~s; median lap 50.15~s) neither the camera nor the 60-GHz branch stopped
the robot and no branch became stale.

\section{Discussion and Conclusion}
\label{sec:conclusion}

\textit{Limitations.} The policy is a supervisory layer over REST and ROS~2
state exchange without timing guarantees; it is not a certified safety
function in the sense of ISO~3691-4. The 60-GHz branch detects people that
alter the propagation along the link rather than persons anywhere in the room,
and its model is site-specific: a change of link geometry required new data
and retraining, and an empty reference recorded with the room as it is
operated. Path-aware gating is only as good as the configured supports
$\Omega_s$ and the representative points $p_s$; an under-estimated support
converts a localization error into a missed hazard. The online runs are
small (6--10 repetitions per condition, one site, open-loop patrol), the
discounting of self-explained RF evidence was not exercised because the patrol
did not cross the link, and the tolerant mode left the 60-GHz branch alone
with two false stops during a camera outage.

\textit{Conclusion.} Heterogeneous infrastructure and robot-borne sensing
becomes useful for robot operation when the DT relates each observation to
where it is valid and to where the robot is going. A commodity 60-GHz link
carrying near-gigabit traffic detected people through the phase signature of
its array, a result that only session-disjoint evaluation reveals, but it does
not transfer between sites without retraining. On a quadruped patrolling
under the DT, every entry into the camera field or the 60-GHz link stopped the
robot, within 1.8~s of physical entry for the camera, a figure dominated by the
confirmation window and braking rather than by sensing, the DT, or the
network. Evaluating the evidence against the corridor the robot is about to
sweep removed every stop caused by off-path presence while still stopping
whenever the corridor reached a person, and the freshness of each branch made the reaction to a
lost branch an explicit, configurable choice. Future work will select among
waiting, slowing down, and re-routing with a cost model over the evidence held
in the DT, and will localize the RF evidence instead of relying on configured
supports.

\ifanon\else
\section*{Acknowledgment}
This work has been partially funded by the European Commission Horizon Europe
SNS JU MultiX (GA 101192521) Project, and by the Regional Government of
Comunidad de Madrid under grant agreement No.\ TEC-2024/COM-360 (Disco6G); and
it is part of the Project 6GINSPIRE PID2022-137329OB-C42, funded by
MCIN/AEI/10.13039/501100011033/ and TEST-6G, funded by TSI-064100-2023-12. We
would like to thank SLICES-Madrid for their services
(\url{https://www.slices-madrid.eu/funding-acknowledgements/}).
\fi

\balance
\bibliographystyle{IEEEtran}
\bibliography{references}
\end{document}